\pdfoutput=1

\documentclass[11pt]{article}

\usepackage[]{emnlp2021}

\usepackage{times}
\usepackage{latexsym}
\usepackage{microtype}
\usepackage{graphicx}
\usepackage{booktabs}  
\usepackage{multirow}
\usepackage{subfigure}
\usepackage{verbatim}
\usepackage{CJKutf8} 
\usepackage{amssymb}
\usepackage{amsmath}

\usepackage{algorithm}
\usepackage{algorithmic}
\usepackage{makecell}
\usepackage{flushend}
\usepackage[T1]{fontenc}

\usepackage[utf8]{inputenc}

\usepackage{microtype}

\title{Quality-aware News Recommendation}

\author{Chuhan Wu$^\dagger$~~~~Fangzhao Wu$^\ddagger$~~~~Tao Qi$^\dagger$~~~~\textbf{Yongfeng Huang}$^\dagger$\\
    $^\dagger$Department of Electronic Engineering, Tsinghua University, Beijing 100084, China  \\
     $^\ddagger$Microsoft Research Asia, Beijing 100080, China\\
  \tt{\{wuchuhan15,wufangzhao,taoqi.qt\}@gmail.com,}\\
  \tt{yfhuang@tsinghua.edu.cn}
  }

\begin{document}
\maketitle
\begin{abstract}

News recommendation is a core technique used by many online news platforms.
Recommending high-quality news to users is important for keeping good user experiences and news platforms' reputations.
However, existing news recommendation methods mainly aim to optimize news clicks while ignore the quality of news they recommended, which may lead to recommending  news with uninformative content or even clickbaits.
In this paper, we propose a quality-aware news recommendation method named \textit{QualityRec} that can effectively improve the quality of recommended news.
In our approach, we first propose an effective news quality evaluation method based on the distributions of users' reading dwell time on news.
Next, we propose to incorporate news quality information into user interest modeling by designing a content-quality attention network to select clicked news based on both news semantics and qualities.
We further train the recommendation model with an auxiliary news quality prediction task to learn quality-aware recommendation model, and we add a  recommendation quality regularization loss to encourage the model to recommend higher-quality news.
Extensive experiments on two real-world datasets show that \textit{QualityRec} can effectively improve the overall quality of recommended news and reduce the recommendation of low-quality news, with even slightly better recommendation accuracy.

\end{abstract}

\section{Introduction}

News recommendation is important for online news platforms to provide users with personalized news reading services to alleviate their information overload~\cite{wu2020mind}.
Predicting whether a user will click on a candidate news is a core task in news recommendation~\cite{wu2021personalized}.
For example, \citet{okura2017embedding} proposed to use a GRU network to model user interest from clicked news, and match it with candidate news for click prediction.
\citet{wu2019nrms} proposed to use multi-head self-attention networks to model the contexts of click behaviors for user modeling, and evaluate the relevance between user interest and candidate news for click prediction.
\citet{wang2020fine} proposed a fine-grained interest matching method based on a 3-D CNN model to predict news click scores.
These methods are dedicated to optimizing news clicks while ignoring the quality of news they recommend.
In fact,  there are inevitably some low-quality news on  news platforms that are not detected by human editors or automatic quality check systems~\cite{shu2018deep}.
These news may have uninformative and even harmful content, and frequently recommending them will heavily injure user experience and the reputation of online news platforms~\cite{urban2014news}.

\begin{figure}[!t]
  \centering 
      \includegraphics[width=0.99\linewidth]{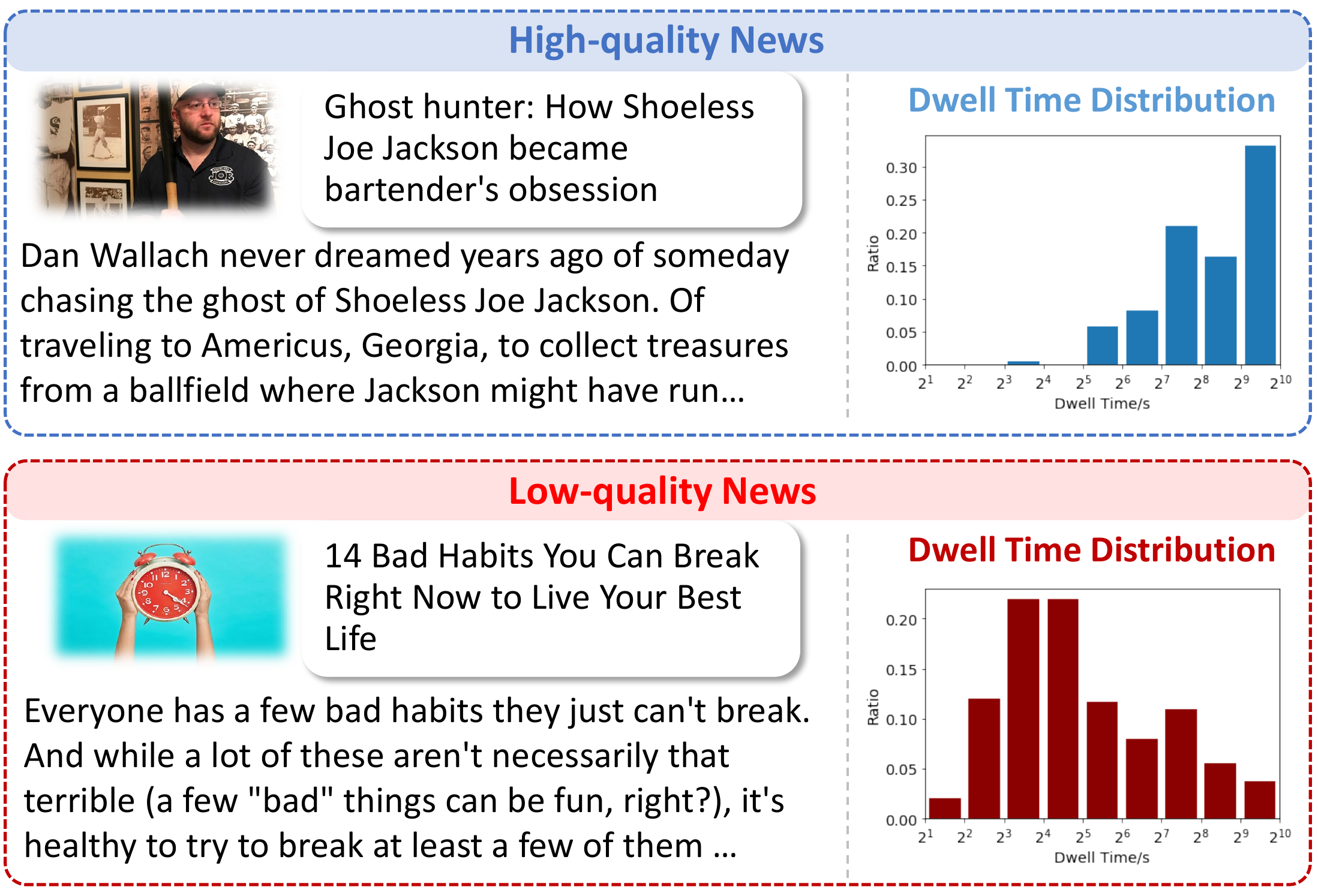} 
  \caption{Examples of a high-quality news and a low-quality news with their dwell time distributions.}\label{fig.exp}  
\end{figure}

To address the above challenge, in this work we study quality-aware news recommendation.
Instead of using manually annotated quality ratings to evaluate news quality, we observe that users' implicit feedback like the dwell time of news reading can be good indications of news quality.
Fig.~\ref{fig.exp} shows an example of a high-quality news and a low-quality news as well as the distributions of users' dwell time on them.
We can see that the high-quality news has a formal and informative title as well as detailed information of the baseball player in the news body.
By contrast, the title of the low-quality news is written in a clickbaity style and the body is also informal and less informative.
We can see that users' dwell time on the high-quality news is usually long and few users leave the news webpage quickly after click, which means that most users are attracted by the content of this news after they click it.
In contrast, most users close the webpage of the low-quality news very quickly, indicating that they are disappointed at its content.
Thus, the distribution of dwell time derived from users' implicit feedback can provide abundant and accessible clues for measuring news  quality.

In this paper, we propose a news quality-aware news recommendation method named \textit{QualityRec}, which can consider the quality factor of news to enhance news recommendation.
In our approach, we first propose an effective method to measure news quality based on the distributions of users' dwell time on each news.
We then design a content-quality joint attention network to consider both semantic and quality information of news when selecting important click behaviors to model user interest in a quality-aware way.
Finally, we propose to train the model with an auxiliary news quality prediction task to learn a quality-aware news model,  and we further introduce an additional regularization function to encourage the model to recommend news with high qualities.
Extensive experiments on two real-world datasets show that \textit{QualityRec} can effectively improve the average recommendation quality and greatly reduce the recommendation chances of low-quality news, with even a slightly better recommendation accuracy.

The contributions of this paper are as follows:
\begin{itemize}
    \item To our best knowledge, this is the first work that studies the problem of quality-aware news recommendation.
    \item We propose an effective method to measure news quality based on users' implicit dwell time feedback.
    \item We design a content-quality attention network to consider both  semantic and quality information of news in user interest modeling.
    \item We develop a quality-aware model training framework with an auxiliary news quality prediction task and a quality regularizer to learn quality-enhanced recommendation models.
\end{itemize}

\section{Related Work}

News recommendation techniques have been extensively studied over years~\cite{wu2021personalized}.
Most existing news recommendation methods rely on click signals to model user interest and train recommendation models~\cite{okura2017embedding,wang2018dkn,wu2019nrms,wang2020fine,liu2020kred,hu2020graph1,hu2020graph,qi2021kim,zhang2021unbert,zhang2021amm}.
For example, \citet{okura2017embedding} proposed to use a GRU network to learn user interest embedding from clicked news, and use the inner product between user embedding and news embedding to predict click scores.
\citet{wang2018dkn} used a candidate-aware attention network to select clicked news according to their relevance to candidate news for user modeling, and compute click scores from the concatenation of news and user embeddings.
\citet{wu2019nrms} proposed to use multi-head self-attention networks to capture the relatedness between clicked news for user modeling, and evaluate the inner product between user and news embeddings for click prediction.
\citet{wang2020fine} explored using a 3-D CNN network to match clicked news and candidate news for click prediction based on their fine-grained semantic relations. 
\citet{qi2021pprec} proposed a popularity-aware news recommendation method that considers popularity information of clicked news and candidate news in both user modeling and click prediction.
However, these methods can only optimize news clicks while ignore the quality of news they recommend, which may lead to recommending some low-quality news such as clickbaits that attract users to click.

\begin{figure*}[!t]
  \centering 
      \includegraphics[width=0.85\linewidth]{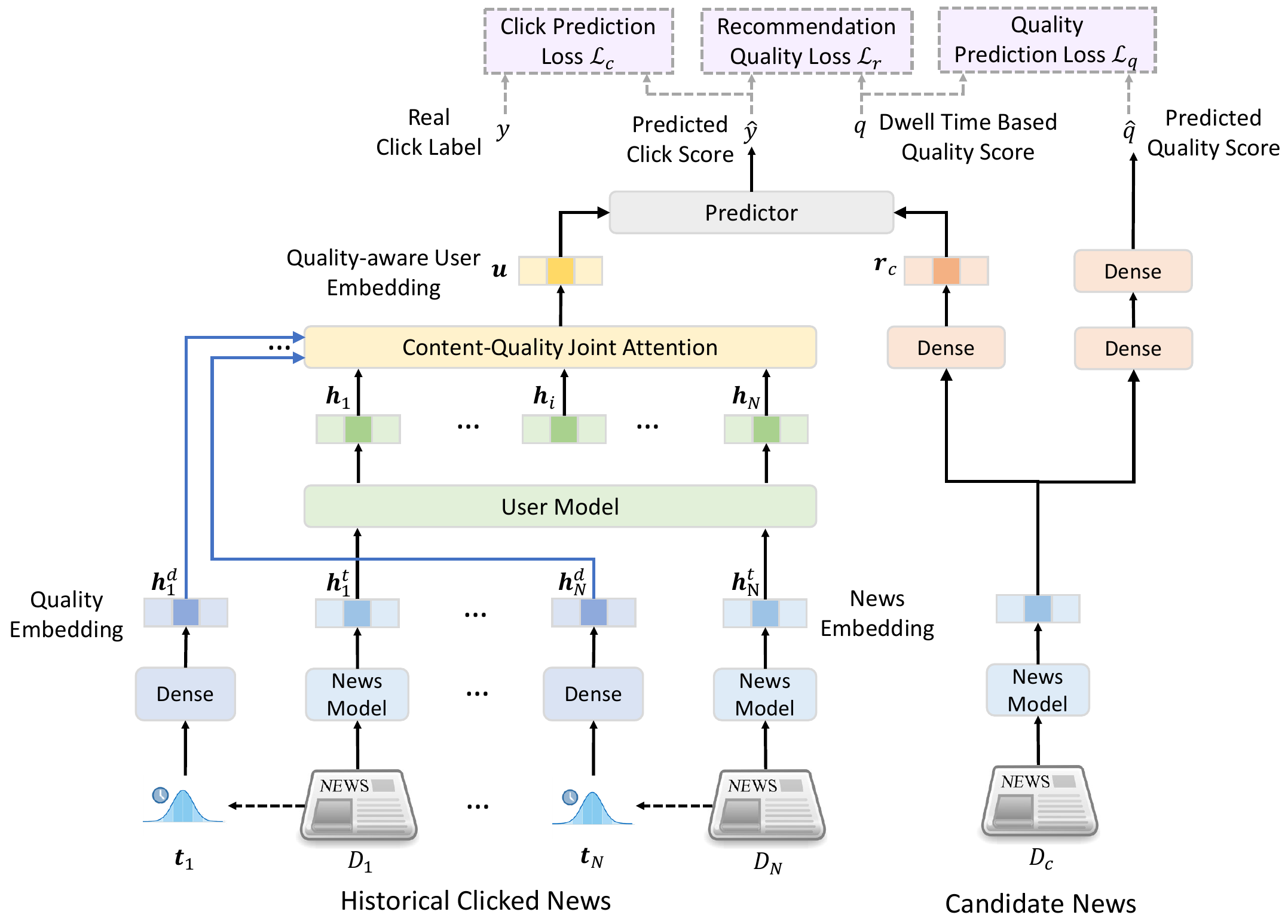} 
  \caption{The framework of \textit{QualityRec}.}\label{fig.model}  
\end{figure*}

There are a few methods that consider user engagement signals beyond clicks to improve news recommendation~\cite{yi2014beyond,wu2020neural,wu2020cprs}.
For example, \citet{yi2014beyond} proposed to weight a user's clicked news in interest modeling based on this user's reading dwell time on them.
\citet{wu2020cprs} proposed a news recommendation method based on click preference and reading satisfaction derived from the normalized reading speed of users to model user interest and train the recommendation models.
However, these methods only consider each individual user's feedback on a news, which cannot effectively indicate the quality of news.
Thus, it is difficult for these methods to surpass the recommendation of low-quality news.
Different from existing methods, we propose an effective method to measure news quality based on implicit user feedback.
We propose to incorporate news quality information into both user modeling and model training, which can effectively promote high-quality news and surpass low-quality news in the recommendation results.

\section{QualityRec}\label{sec:Model}

In this section, we introduce the details of our proposed \textit{QualityRec} method for quality-aware news recommendation, as shown in Fig.~\ref{fig.model}.
We will first introduce how to model news quality with users' implicit feedback, then explain how to incorporate quality information into user modeling, and finally introduce how to train a quality-aware news recommendation model.

\subsection{News Quality Modeling}

Effective modeling of news quality is a prerequisite for quality-aware news recommendation.
However, it is very challenging due to the lack of labeled data and the diverse and dynamic characteristics of online news information~\cite{louis2013makes,wu2020mind}.
Prior work has found that the implicit feedback of users on news such as dwell time can be indications of user satisfaction on news content~\cite{lu2018between,wu2020cprs}, which has the potential to reflect news quality.
However, the satisfaction of each individual user cannot well indicate news quality due to the variety of personal preferences.
An intuitive way is using the average dwell time of users on a specific news as its quality measurement.
However, the average dwell time is heavily influenced by outliers and cannot reflect the ratios of short dwell time, which is usually important for detecting low-quality news~\cite{wu2020neural}.
Thus, average dwell time may be suboptimal in news quality modeling.
To solve this problem, we propose to use the distribution of dwell time on news to measure its quality.
Since the length dwell time may have some randomness, we first quantize each continuous dwell time value $d_i$ into a discrete one with the  logarithm function as follows:
\begin{equation}
    \hat{d}_i=\lfloor\log_2(1+d_i)\rfloor.
\end{equation}
We denote the quantized dwell time of a news as $[\hat{d}_1, \hat{d}_2, ..., \hat{d}_C]$, where $C$ is number of clicks on it.
We represent the dwell time distribution of this news as a vector $\mathbf{t}$, and its $j-th$ element $t_j$ can be computed as follows:
\begin{equation}
    t_j=\frac{1}{C}\sum_k{I(\hat{d}_k=j)},
\end{equation}
where $I(\cdot)$ is an indicator function.
Since the vector $\mathbf{t}$ contains the  distribution information of dwell time on a news, it is more informative than average dwell time in modeling news quality.

In addition, since we observe that the length of dwell time is approximately lognormal, the following formulation\footnote{The element index $i$ starts from 0.} can be used if we need to directly derive a quality score $q$ from $\mathbf{t}$:
\begin{equation}
    q=\sum_i i*t_i,
\end{equation}
which means that the news quality score is higher if a higher ratio of clicks is associated with longer dwell time.
In this way, we can obtain a simple yet effective method for measuring news quality from implicit user feedback.

\subsection{Quality-aware User Modeling}

We then introduce the quality-aware user modeling method in \textit{QualityRec}.
Existing user modeling methods usually only consider the feedback of each individual user into interest modeling, while the influence of the quality of clicked news on user interest modeling and subsequent news recommendation cannot be fully considered.
Thus, in our approach we explore to consider quality information in user modeling.
We denote the historical clicked news of a user as $[D_1, D_2, ..., D_N]$, where $N$ is the history length.
Based on our proposed quality modeling method, we first obtain the dwell time distributions of clicked news, which are denoted as $[\mathbf{t}_1, \mathbf{t}_2, ..., \mathbf{t}_N]$.
We use a dense layer to learn hidden representations of them (denoted as $[\mathbf{h}^d_1, \mathbf{h}^d_2, ..., \mathbf{h}^d_N]$) to better capture quality information.
To model the content information of clicked news, following prior work~\cite{wu2019nrms} we use a news model to encode the titles of clicked news into their hidden semantic representations, which are denoted as $[\mathbf{h}^t_1, \mathbf{h}^t_2, ..., \mathbf{h}^t_N]$.
Without loss of generality, we implement the news model with a Transformer~\cite{vaswani2017attention} encoder with the attention pooling mechanism.
We then use a user model to capture the relatedness between user behaviors for user interest modeling, and it outputs a sequence of contextualized click behavior representations $[\mathbf{h}_1, \mathbf{h}_2, ..., \mathbf{h}_N]$.
We also implement the user model with a Transformer.
Finally, to select both informative and qualified news for user interest modeling, we propose a content-quality joint attention network to pay different attention to clicked news according to both their semantic and quality information.
More specifically, the content attention score $\alpha^c_i$ and the quality attention score $\alpha^q_i$ of the $i$-th clicked news are computed as follows:
\begin{equation}
    \alpha^c_i=\mathbf{v}^c\tanh(\mathbf{W}^c\mathbf{h}_i+\mathbf{b}^c),
\end{equation}
\begin{equation}
    \alpha^q_i=\mathbf{v}^q\tanh(\mathbf{W}^q\mathbf{h}_i+\mathbf{b}^q),
\end{equation}
where $\mathbf{v}^c$, $\mathbf{v}^q$, $\mathbf{W}^c$, $\mathbf{W}^q$, $\mathbf{b}^c$, and $\mathbf{b}^q$ are model parameters.
The normalized content-quality joint  attention score $\alpha_i$ is formulated as follows:
\begin{equation}
    \alpha_i=\frac{\exp(\alpha^c_i+\alpha^q_i)}{\sum_{j=1}^N \exp(\alpha^c_j+\alpha^q_j)}.
\end{equation}
The final quality-aware user interest embedding $\mathbf{u}$ is a weighted summation of contextualized click behavior representations, i.e., $\mathbf{u}=\sum_{i=1}^N \alpha_i \mathbf{h}_i$.
The user embedding is further used for predicting a click score $\hat{y}$ based on its relevance to the candidate news embedding $\mathbf{r}^c$, which is learned from the candidate news title\footnote{We only use news titles for click prediction because users cannot see the main content before click.} through the news encoder and an additional dense layer.
The click score is finally used for personalized news ranking.

\subsection{Model Training}

We then introduce the quality-aware training framework  of \textit{QualityRec}.
The main task is the click prediction task.
Following many existing works~\cite{wu2019npa} we use the negative sampling method to construct training samples  from impressions, and we use crossentropy loss (denoted as $\mathcal{L}_c$) to train the click prediction model.
However, the click prediction task is not aware of news quality signals, and therefore we introduce two additional loss functions to train the model.

The first one is a quality prediction loss, which aims to predict the quality score $q$ of a candidate news derived from its dwell time distribution.
Since the quality of news depends on both its title and main body, we apply the news encoder to the concatenation of news title and body to learn a hidden news representation $\mathbf{h}^c$.
We apply a dense layer to learn quality prediction task-specific news representations from $\mathbf{h}^c$, and use another dense layer to predict the quality score $\hat{q}$.
To train the quality prediction model, we use the mean absolute error as the loss function (denoted as $\mathcal{L}_q$), which is formulated as follows:
\begin{equation}
    \mathcal{L}_q = \frac{1}{Q}|\hat{q}-q|,
\end{equation}
where $Q$ is the maximum value of $q$ in the training dataset (for normalization purpose).\footnote{We use the quality score as the prediction target rather than the raw dwell time distributions because we find it is quite difficult to predict dwell time distributions from news content due to the uncertainty of user behaviors.}

The second one is a recommendation quality regularization.
It encourages the  model to give high click scores for high-quality news and demote low-quality news.
To achieve this goal, we design a regularization function $\mathcal{L}_r$ as follows:
\begin{equation}
    \mathcal{L}_r = \frac{\hat{y}}{\epsilon+q},
\end{equation}
where $\epsilon$ is a small value (i.e., 1e-6) to avoid the click score being divided by a  zero quality score.
In this loss, if the quality of top ranked news is lower, the regularization loss will be larger.
Thus, the model will be encouraged to recommend news with higher quality by optimizing this loss.
The unified loss $\mathcal{L}$ for joint model training is a weighted summation of the above three kinds of losses:
\begin{equation}
    \mathcal{L}=\mathcal{L}_c+\lambda \mathcal{L}_q+ \mu \mathcal{L}_r,
\end{equation}
where $\lambda$ and  $\mu$ are hyperparameters that control the intensity of the two auxiliary losses.

\section{Experiments}\label{sec:Experiments}

\subsection{Datasets and Experimental Settings}

\begin{figure}[!t]
	\centering 
	\subfigure[\textit{News}.]{
	\includegraphics[width=0.4\textwidth]{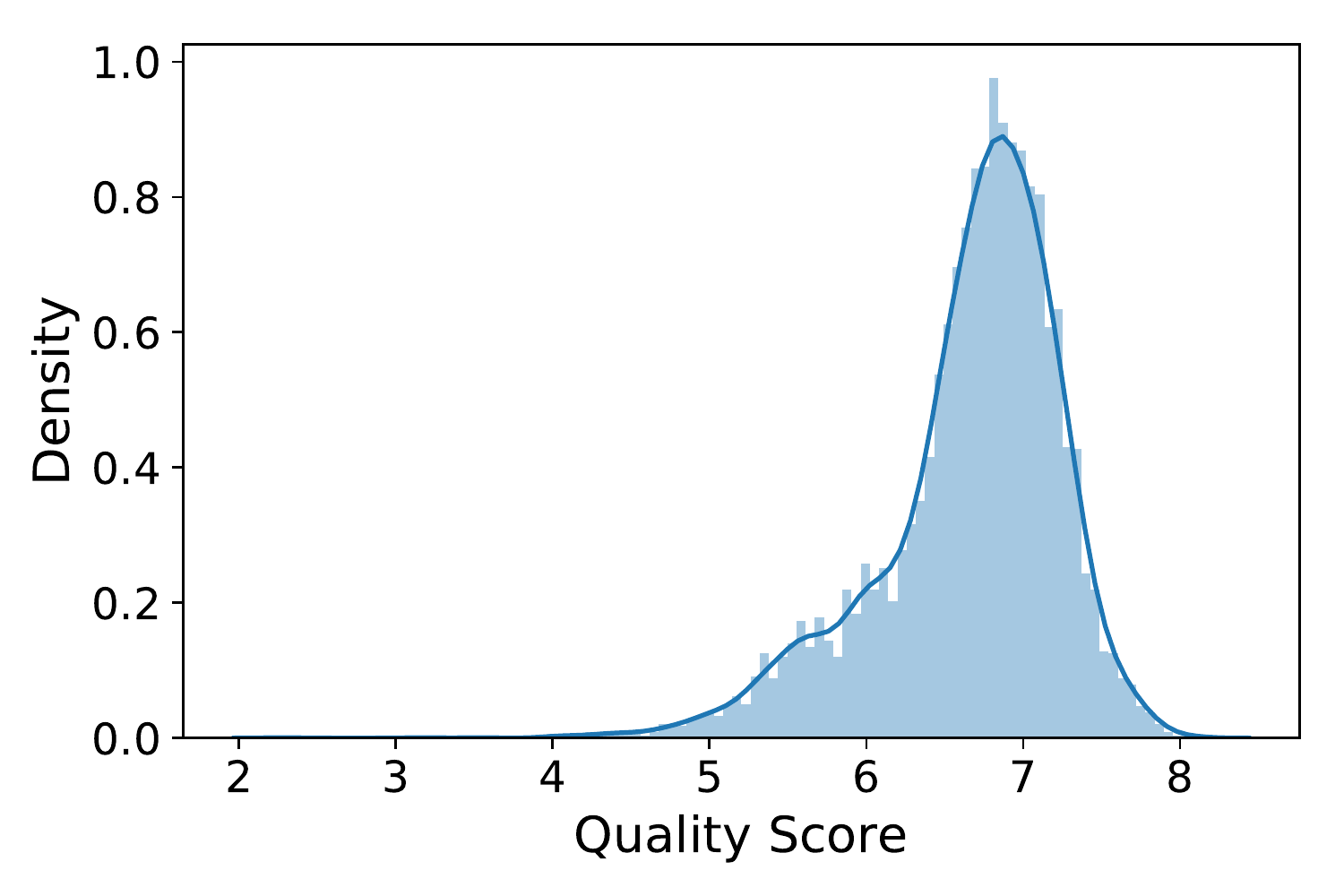} 
	}
		\subfigure[\textit{Feeds}.]{
	\includegraphics[width=0.4\textwidth]{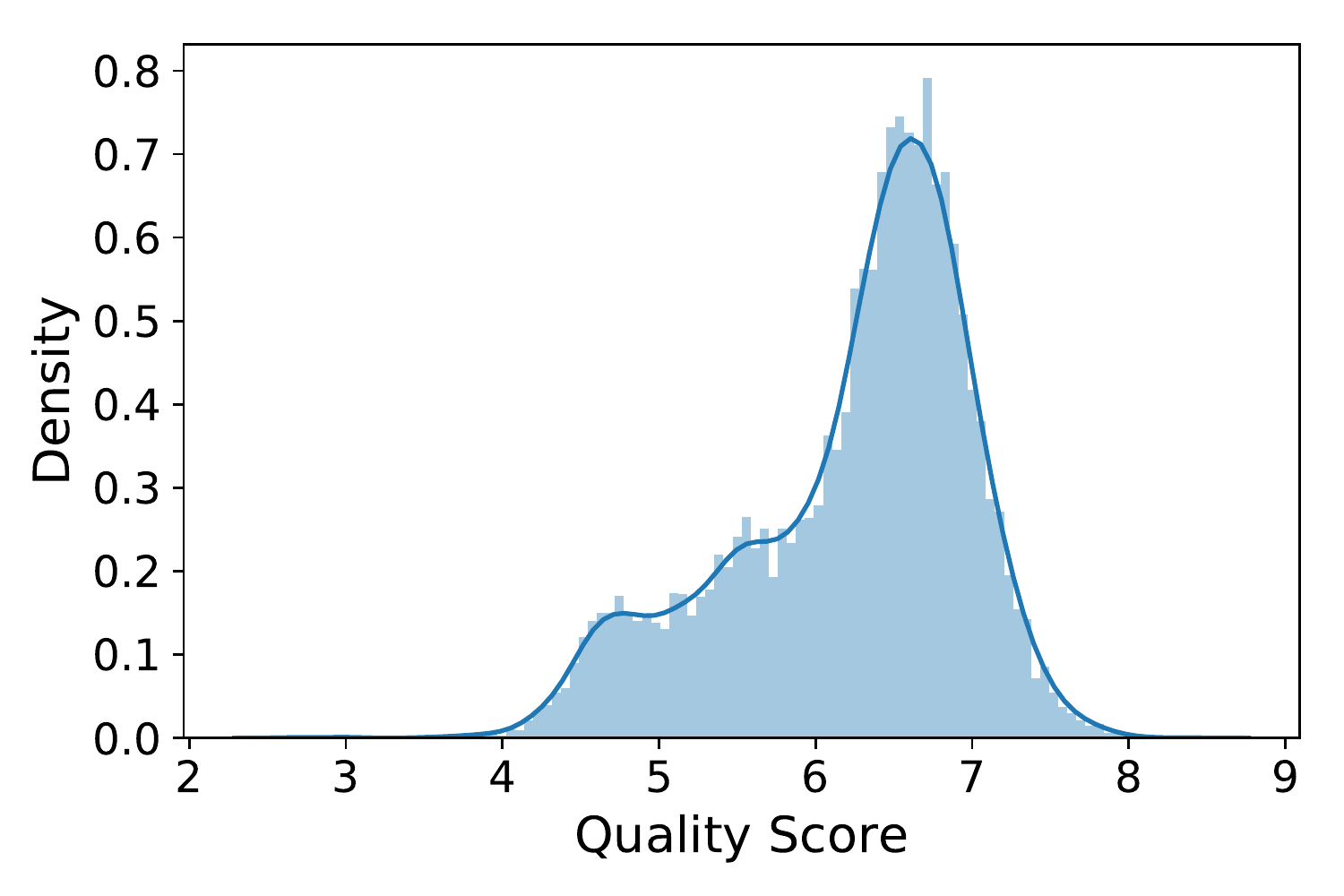} 
	}
\caption{Distributions of quality scores.}\label{fig.qua}
\end{figure}

We use two real-world news recommendation datasets for conducting our experiments.
The first one is drawn from~\cite{wu2020cprs}, which is a dataset collected from Microsoft News (denoted as \textit{News}).
It contains 500,000 news impressions of 285,563 users from 10/12/2019 to 11/13/2019.
The second one is a proprietary dataset collected by our self from a commercial news feeds platform  (denoted as \textit{Feeds}.\footnote{Anonymized for double-blind review.}
It  contains 500,000 impression logs of 337,286 users between 06/23/2019 and 07/20/2019.
On both datasets, the impressions in the last week are used for test, and the rest are used for training and validation.
More detailed statistical information of the two datasets is shown in Table~\ref{table.dataset}.
In addition, we show the distributions of quality score on both datasets in Fig.~\ref{fig.qua}.

\begin{table}[h]
\centering
\begin{tabular}{lrr}
\Xhline{1.0pt}
                   & \multicolumn{1}{c}{\textbf{News}} & \multicolumn{1}{c}{\textbf{Feeds}} \\ \hline
\#news             & 95,628                   & 112,052                   \\
\#users            & 285,563                  & 337,286                   \\
\#impressions      & 500,000                  & 500,000                   \\
\#clicks           & 790,588                  & 853,291                   \\
avg. title len.    & 11.65                    & 11.28                     \\
avg. dwell time    & 293.17                   & 267.98                    \\
avg. quality score & 6.719                    & 6.297                     \\ \Xhline{1.0pt}
\end{tabular}
	\caption{Statistical information of datasets.}\label{table.dataset}
\end{table}

 \begin{table*}[!t]
\centering
\resizebox{0.999\linewidth}{!}{ 
\begin{tabular}{lcccccccc}
\Xhline{1.0pt}  
\multicolumn{1}{c}{\textbf{Methods}} & AUC($\uparrow$)    & MRR($\uparrow$)    & nDCG@5($\uparrow$) & nDCG@10($\uparrow$) & QS@5($\uparrow$)  & QS@10($\uparrow$) & LQ@5($\downarrow$)    & LQ@10($\downarrow$)  \\ \hline
EBNR                                 & 0.6086          & 0.2025          & 0.2068          & 0.2531          & 6.757          & 6.751          & 0.00638          & 0.0137          \\
DKN                                  & 0.6020          & 0.1981          & 0.2001          & 0.2479          & 6.760          & 6.755          & 0.00634          & 0.0136          \\
NAML                                 & 0.6192          & 0.2093          & 0.2133          & 0.2616          & 6.764          & 6.757          & 0.00632          & 0.0135          \\
NPA                                  & 0.6223          & 0.2114          & 0.2150          & 0.2644          & 6.758          & 6.750          & 0.00640          & 0.0137          \\
NRMS                                 & 0.6263          & 0.2140          & 0.2182          & 0.2673          & 6.761          & 6.755          & 0.00636          & 0.0136          \\
FIM                                  & 0.6301          & 0.2167          & 0.2211          & 0.2709          & 6.756          & 6.748          & 0.00642          & 0.0139          \\
HieRec                               & 0.6376          & 0.2198          & 0.2275          & 0.2771          & 6.732          & 6.724          & 0.00649          & 0.0141          \\ \hline
DTW                                  & 0.6310          & 0.2135          & 0.2220          & 0.2718          & 6.772          & 6.764          & 0.00624          & 0.0134          \\
CPRS                                 & 0.6475          & 0.2231          & 0.2321          & 0.2830          & 6.781          & 6.770          & 0.00616          & 0.0131          \\ \hline
QualityRec                           & \textbf{0.6481} & \textbf{0.2235} & \textbf{0.2323} & \textbf{0.2834} & \textbf{6.844} & \textbf{6.841} & \textbf{0.00460} & \textbf{0.0109} \\ \Xhline{1.0pt}
\end{tabular} 
}
\caption{Recommendation accuracy and quality on \textit{News}. ``$\uparrow$'': higher is better.  ``$\downarrow$'': lower is better.}  \label{p1}
\label{result}
\end{table*}

 \begin{table*}[!t]
\centering
\resizebox{0.999\linewidth}{!}{ 
\begin{tabular}{lcccccccc}
\Xhline{1.0pt}
\multicolumn{1}{c}{\textbf{Methods}} & AUC($\uparrow$)    & MRR($\uparrow$)    & nDCG@5($\uparrow$) & nDCG@10($\uparrow$) & QS@5($\uparrow$)  & QS@10($\uparrow$) & LQ@5($\downarrow$)    & LQ@10($\downarrow$)  \\ \hline
EBNR                                 & 0.6564          & 0.2443          & 0.2650          & 0.3246          & 6.593          & 6.576          & 0.00198          & 0.00469          \\
DKN                                  & 0.6528          & 0.2395          & 0.2606          & 0.3208          & 6.585          & 6.572          & 0.00201          & 0.00474          \\
NAML                                 & 0.6672          & 0.2498          & 0.2702          & 0.3279          & 6.606          & 6.588          & 0.00186          & 0.00444          \\
NPA                                  & 0.6680          & 0.2501          & 0.2706          & 0.3284          & 6.604          & 6.587          & 0.00189          & 0.00448          \\
NRMS                                 & 0.6696          & 0.2511          & 0.2721          & 0.3294          & 6.607          & 6.590          & 0.00188          & 0.00446          \\
FIM                                  & 0.6714          & 0.2524          & 0.2735          & 0.3319          & 6.602          & 6.585          & 0.00191          & 0.00452          \\
HieRec                               & 0.6765          & 0.2565          & 0.2770          & 0.3359          & 6.599          & 6.581          & 0.00194          & 0.00461          \\ \hline
DTW                                  & 0.6704          & 0.2518          & 0.2739          & 0.3301          & 6.611          & 6.594          & 0.00185          & 0.00439          \\
CPRS                                 & 0.6787          & 0.2582          & 0.2798          & 0.3375          & 6.620          & 6.605          & 0.00179          & 0.00416          \\ \hline
QualityRec                           & \textbf{0.6801} & \textbf{0.2594} & \textbf{0.2816} & \textbf{0.3389} & \textbf{6.696} & \textbf{6.688} & \textbf{0.00134} & \textbf{0.00318} \\ 
 \Xhline{1.0pt}
\end{tabular} 
}
\caption{Recommendation accuracy and quality on \textit{Feeds}. ``$\uparrow$'': higher is better.  ``$\downarrow$'': lower is better.}  \label{p2}
\label{result}
\end{table*}

In our experiments, following~\cite{wu2019nrms,wang2020fine} we use GloVe~\cite{pennington2014glove} to initialize word embeddings in the news model.
The hidden dimension in the news and user models is 400, and the Transformers in them have 20 attention heads.
We use Adam~\cite{kingma2014adam} as the model optimizer and the learning rate is 0.0001.
The intensity of dropout applied to each layer is 0.2.
The loss weights $\lambda$ and $\mu$ are 2.0 and 0.5, respectively.\footnote{More details of hyperparameter  are included in Appendix.}
To measure recommendation accuracy, we use AUC, MRR, nDCG@5 and nDCG@10.
To measure the quality of recommended news, we use two  types of metrics.
The first one is the average quality score of top K ranked news (denoted as QS@K).
The second one is the number of news with the lowest 5\% quality score that appear in the top K recommendation results (denoted as LQ@K).
We report the results under K=5 and K=10.
We repeat each experiment 5 times and report the average scores.

\subsection{Performance Evalution}

First, we evaluate the performance of different methods in terms of their recommendation accuracy and the quality of top recommended news (shorten as recommendation quality).
We compare \textit{QualityRec} with the following baseline methods:
(1) \textit{EBNR}~\cite{okura2017embedding}, an embedding-based news recommendation method with a GRU network for user modeling;
(2) \textit{DKN}~\cite{wang2018dkn}, using knowledge-aware CNN for news modeling and candidate-aware attention for user modeling;
(3) \textit{NAML}~\cite{wu2019}, using CNN and attention networks for news modeling and an attention network for user modeling;
(4) \textit{NPA}~\cite{wu2019npa}, using personalized attention mechanism for both news and user modeling;
(5) \textit{NRMS}~\cite{wu2019nrms}, using multi-head self-attention mechanism for news and user modeling;
(6) \textit{FIM}~\cite{wang2020fine}, a fine-grained interest matching method based on 3-D CNN for news recommendation.
(7) \textit{HieRec}~\cite{qi2021hierec}, a hierarchical interest modeling and matching method for news recommendation.
(8) \textit{DTW}~\cite{yi2014beyond}, using dwell time to weight clicked news in user modeling.
We use \textit{NRMS} as the basic model.
(9) \textit{CPRS}~\cite{wu2020cprs}, a user modeling method with click preference and reading satisfaction for news recommendation.
The results on the \textit{News} and \textit{Feeds} datasets are shown in Tables~\ref{p1} and~\ref{p2}, respectively.
We have third main observations from the results.
First, the methods based on click signals only usually have lower recommendation quality than those that consider user engagement signals (i.e., \textit{DTW}, \textit{CPRS}, and \textit{QualityRec}).
This may be because optimizing click-through rate may lead to recommending some news that entice users to click (i.e, clickbaits), which usually have low content quality.
Second, \textit{QualityRec} can more effectively improve recommendation quality than \textit{DTW} and  \textit{CPRS}, and can even slightly boost recommendation accuracy.
This shows that the engagement signals of each individual user are not suitable for indicating news quality, and incorporating quality signals in an appropriate way can also help target user interest.
Third, we find that \textit{QualityRec} is especially effective in reducing the frequency of recommending low-quality news.
It shows that our approach may effectively mitigate the negative influence of low-quality news on user experience.

 \begin{figure}[!t]
	\centering 
	\subfigure[Recommendation accuracy.]{
	\includegraphics[width=0.98\linewidth]{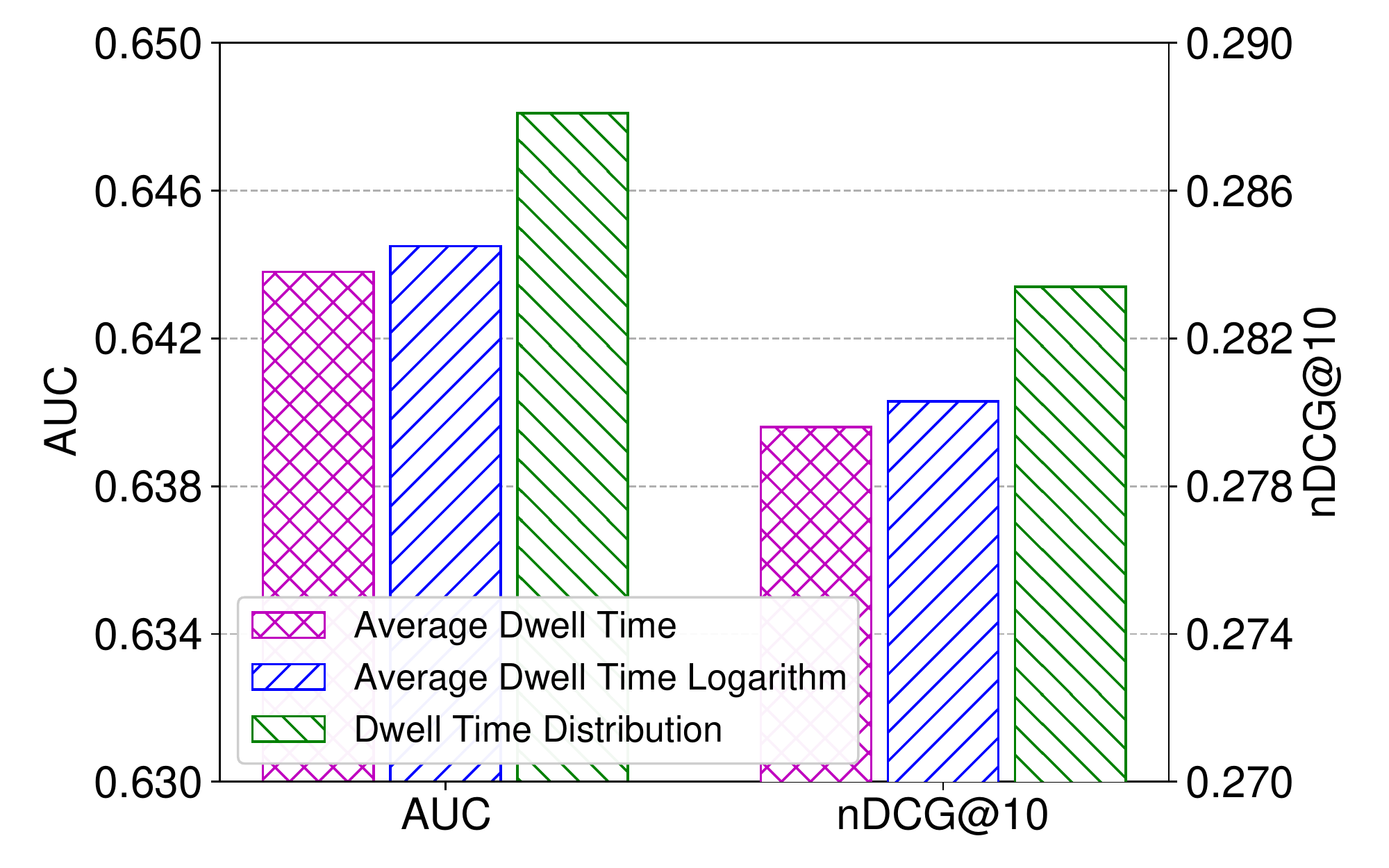} 
	}
	\subfigure[Recommendation quality.]{
	\includegraphics[width=0.98\linewidth]{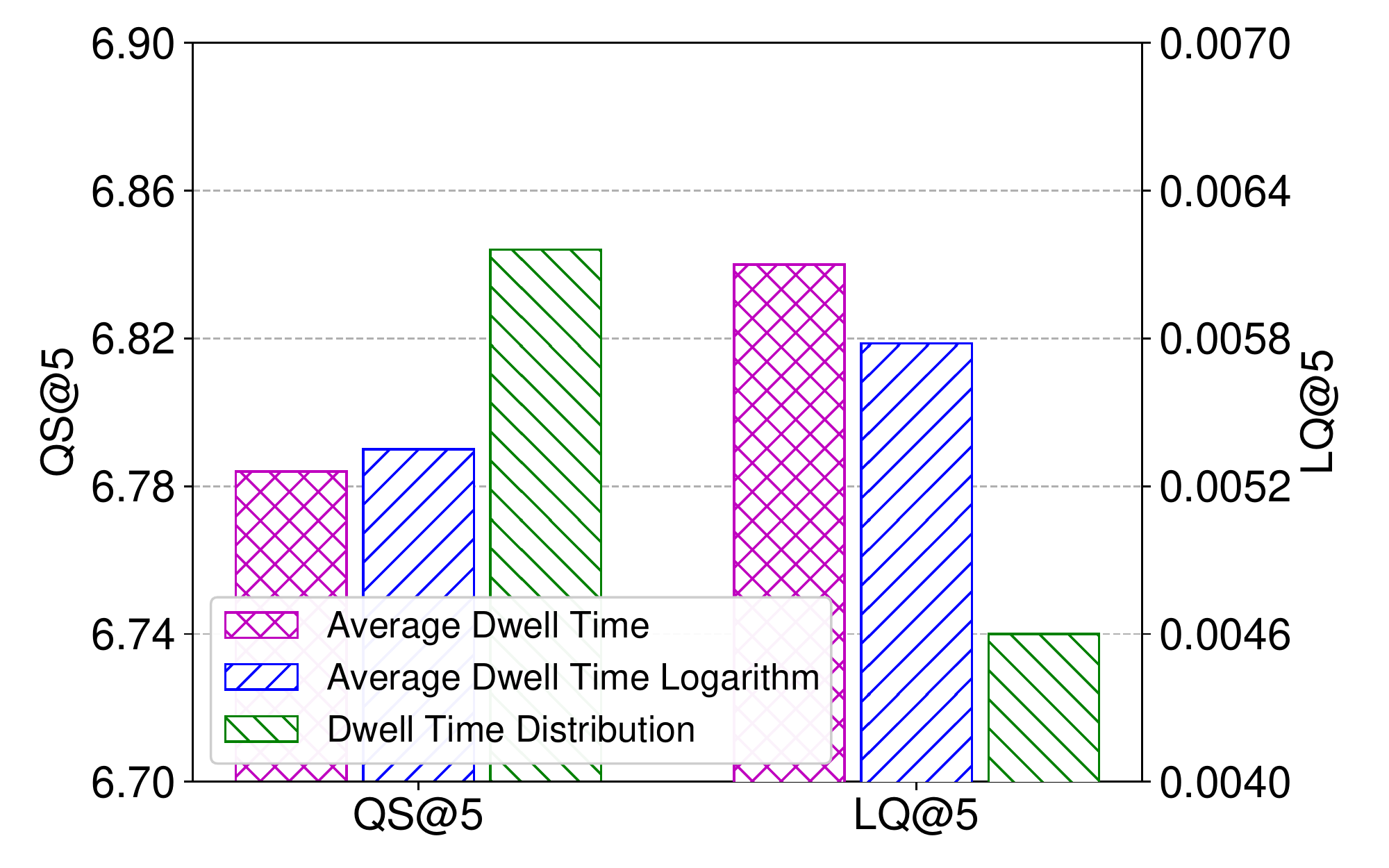} 
	}
\caption{Influence of news quality measurements.}\label{fig.mea}
\end{figure}

\begin{table*}[t]
\resizebox{0.999\linewidth}{!}{ 
\begin{tabular}{ll}
\hline
\multicolumn{1}{c}{\textbf{Low Predicted Quality Score}}         & \multicolumn{1}{c}{\textbf{High Predicted Quality Score}}                                            \\ \hline
Wild \& wonderful treasures of SEMA 2019 – US edition            & The NFL Hot Seat: Sean McVay ran out of magic, and Jared Goff isn't helping                          \\
100 Incredible Christmas Tree Decorating Ideas                   & Hundreds rally at Texas Governor’s Mansion in support of death row inmate Rodney Reed                \\
14 Celebs Over 50 Who Are In The Best Shape Of Their Lives       & Anthony Bourdain's Chef's Knife Sold for Over \$230,000                                              \\
The 22 Wildest Celebrity Hair Transformations of the Last Decade & Free Rodney Reed Petition Surpasses 2 Million Signatures to Stop Execution of Texas Death Row Inmate \\
25 Biggest Grocery Store Mistakes Making You Gain Weight         & Denny Hamlin seals title bid with clinching Phoenix victory                                          \\
Top 30 Mistakes Diyers Make When Remodeling a Kitchen            & 'Beautiful boys': Victims in Mexico ambush remembered as funerals                                    \\
Best Nostalgic Pics of Leonardo Dicaprio                         & Amazon Is Accused of Forcing Up Prices in Antitrust Complaint                                        \\
30 Style Shortcuts for No-hassle Christmas Decor                 & An Arctic blast is coming to the eastern part of the US                                              \\
30 home trends that ruled the last decade                        & Winners and losers in College Football Playoff first rankings include Big Ten and Clemson            \\
These Are The World's Most Beautiful Small Towns                 & Monday Measure: Why Clemson deserves playoff inclusion (and more respect)                            \\ \hline
\end{tabular}
}
\caption{The news with the lowest or the highest predicted quality scores.}\label{fig.score}
\end{table*}

\subsection{Effectiveness of News Quality Modeling}

Next, we verify the effectiveness of the news quality modeling method we proposed.
We first compare the recommendation accuracy and quality under different quality modeling methods, including: (a) using average dwell time of a news as quality measurement; (b) using the average of the logarithm  of dwell time as quality measurement; (c) using the distribution of quantized dwell time.
We conducted experiments on the \textit{News} dataset and the results are shown in Fig.~\ref{fig.mea}.\footnote{The following experimental results are all on \textit{News}.}
We can see that using the average dwell time yields the worst performance.
This is because average dwell time can be heavily influenced by outlier values and can be noisy for quality modeling.
In addition, using the average of dwell time logarithm is also suboptimal.
This is because it also losses the distribution information of dwell time and may be affected by extreme values.
In addition, we use the Pearson correlation coefficient (PCC) to evaluate the quality prediction results under different quality label construction methods, i.e., (a), (b) and the quality score computing method in our approach.
The PCC score of them are 0.22, 0.36 and 0.48, respectively, which shows that the quality scores labeled by our proposed method may be less noisy.\footnote{We provide further discussions on the correlations between them and the quality scores rated by users in Appendix.}

We also show the top 10 news with the lowest or highest quality scores predicted by our approach in Table~\ref{fig.score}.\footnote{The bodies of these news are not shown due to space limit, and they can be retrieved by common search engines.}
We can see that the titles of predicted low-quality news are all written in a clickbaity style, and it is unsuitable to recommend them to users frequently.
By contrast, the predicted high-quality news have attractive and informative titles, which may better meet users' interest. 
This result shows that \textit{QualityRec} not only can serve as a news recommender, but also can be used as a news quality evaluator to help news platforms identify potential high-quality news and filter low-quality ones like clickbaits  for better news quality moderation.

\subsection{Ablation Study}

\begin{figure}[!t]
	\centering 
	\subfigure[Recommendation accuracy.]{
	\includegraphics[width=0.98\linewidth]{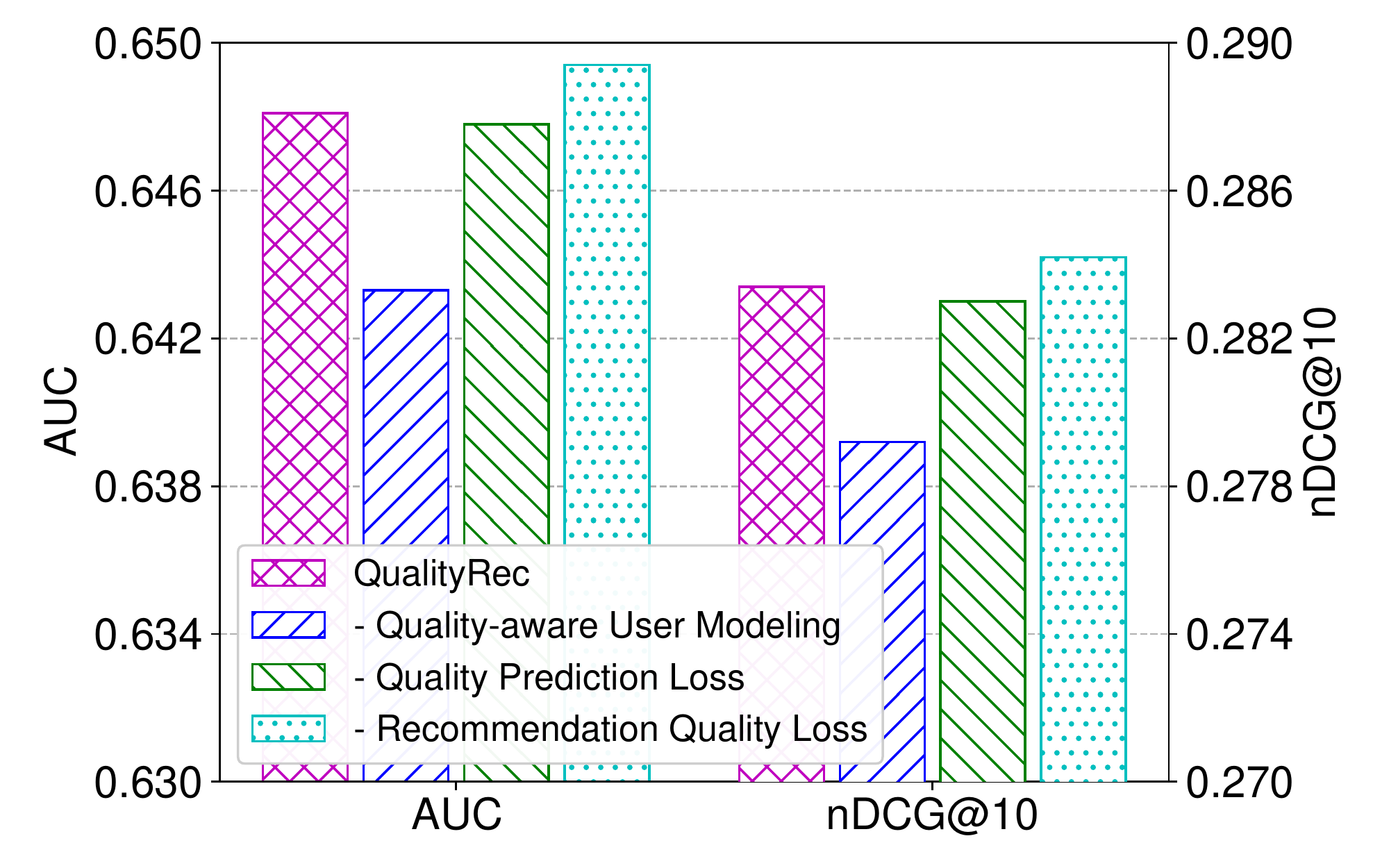} 
	}
	\subfigure[Recommendation quality.]{
	\includegraphics[width=0.98\linewidth]{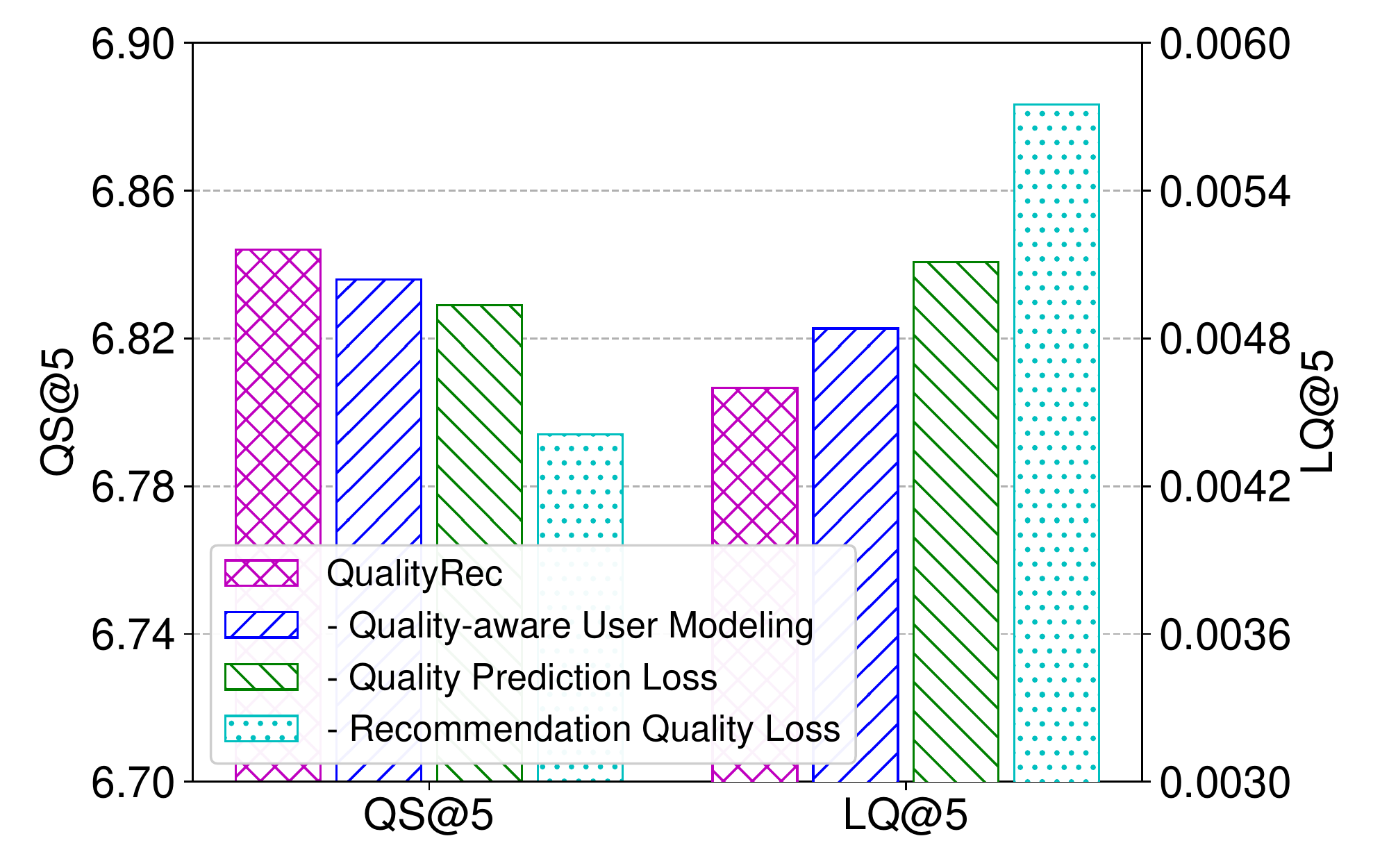} 
	}
\caption{Effectiveness of the quality-aware user modeling technique and different training losses.}\label{fig.ab}
\end{figure}

We then present several ablation studies on the quality-aware user modeling method and the loss functions for model training.
We remove one of them from the model and compare the recommendation accuracy and quality in Fig.~\ref{fig.ab}.\footnote{We remove the quality attention score to deactivate quality-aware user modeling.}
We find that the quality-aware user modeling method can improve both accuracy and quality, which is because quality signals are both useful for targeting user interest and improving recommendation quality.
In addition, the quality prediction loss has notable contributions to recommendation quality, which shows that learning a quality-aware news model can help quality-aware recommendation.
Moreover, we find that the recommendation quality loss leads to some accuracy sacrifices but great quality improvements.
It shows that regularizing the model to recommend high-quality news will lose some clicks, but maybe actually beneficial for user experience.

\subsection{Hyperparameter Analysis}

\begin{figure}[!t]
	\centering 
	\subfigure[Recommendation accuracy.]{
	\includegraphics[width=0.73\linewidth]{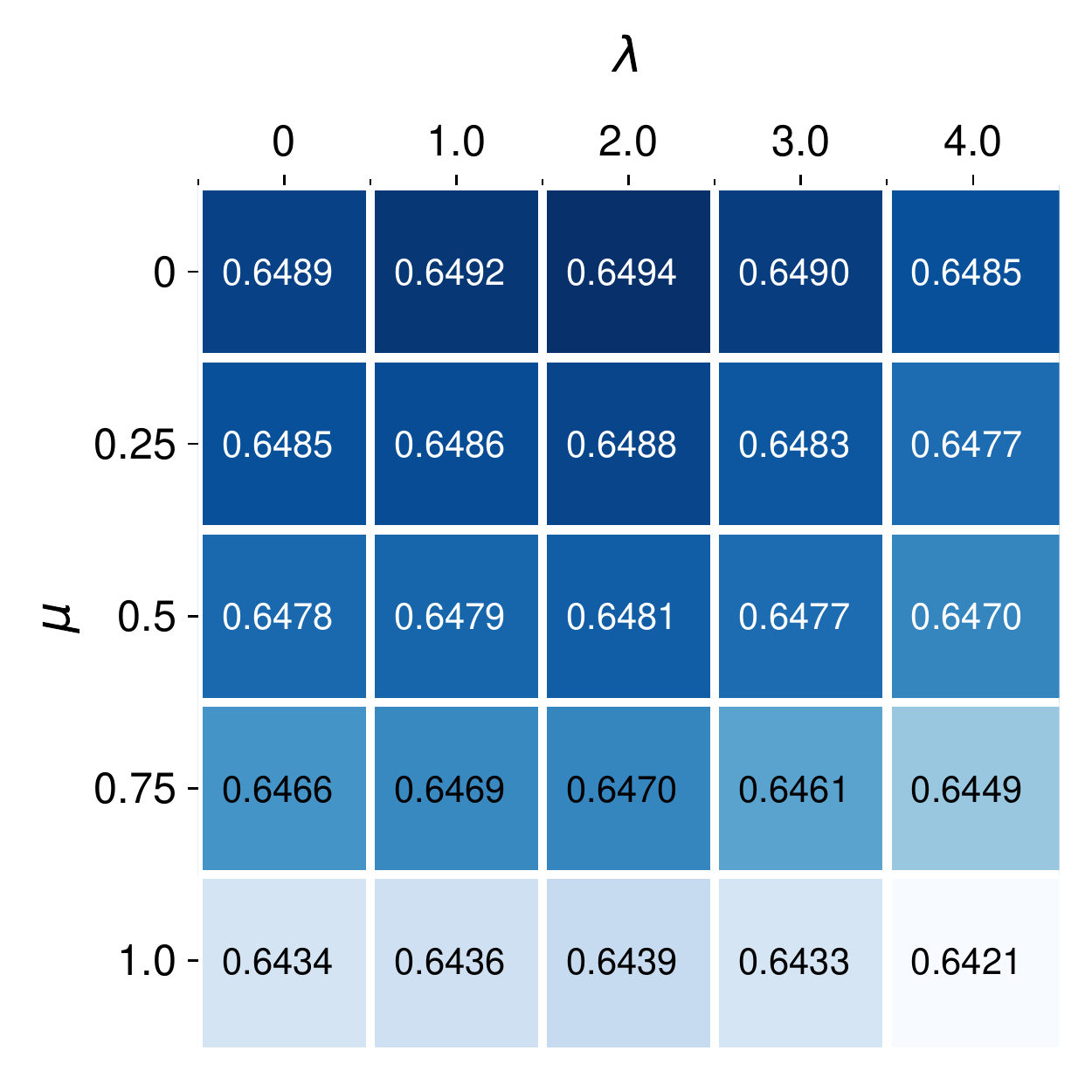} 
	}
	\subfigure[Recommendation quality.]{
	\includegraphics[width=0.73\linewidth]{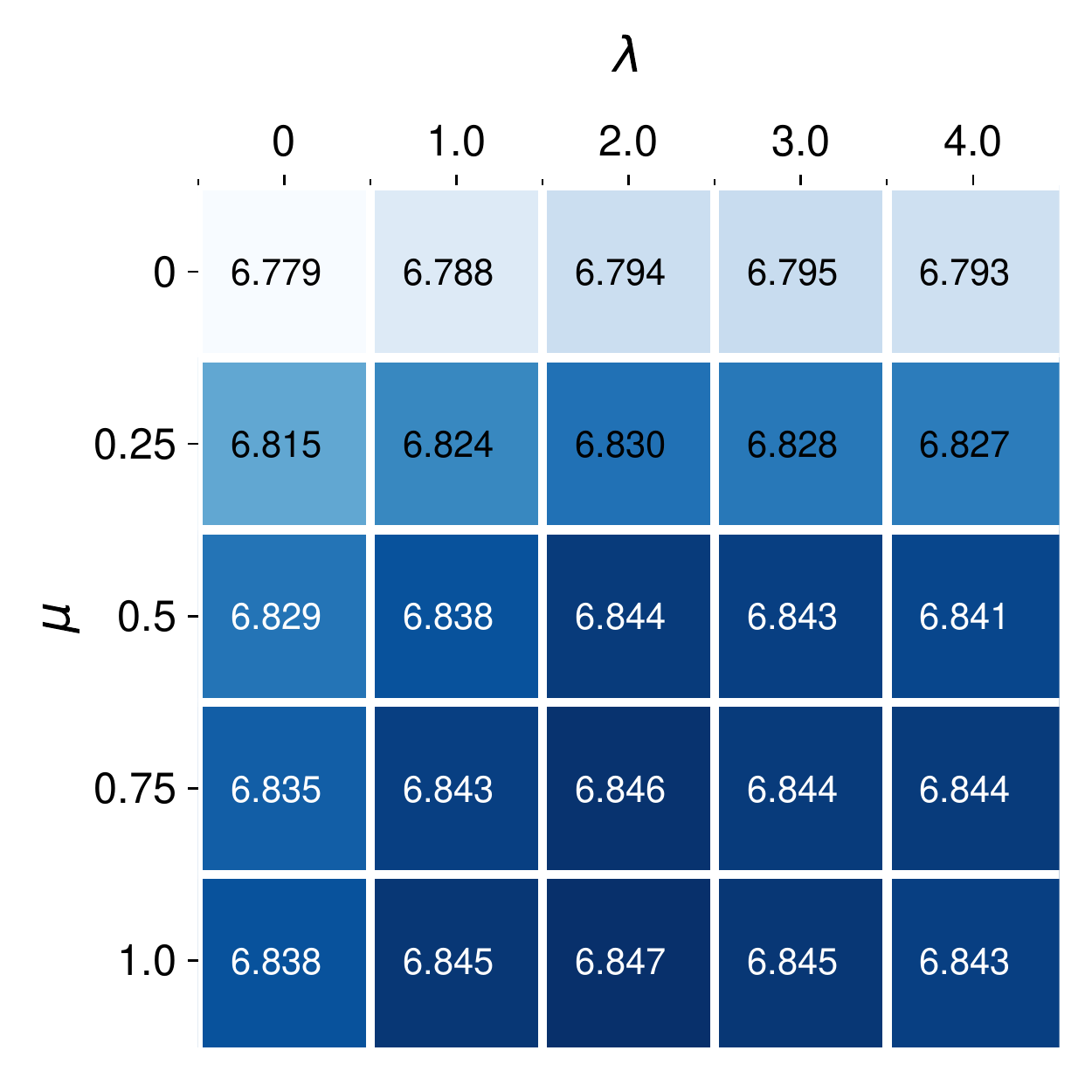} 
	}
\caption{Influence of the hyperparameters $\lambda$ and $\mu$.}\label{fig.hyper}
\end{figure}

Finally, we discuss the influence of the two loss coefficients (i.e., $\lambda$ and $\mu$) in Eq. (9).
We vary the quality prediction loss weight $\lambda$ from 0 to 4.0 and the recommendation quality loss weight $\mu$ from 0 to 1.
The results are shown in Fig.~\ref{fig.hyper}.
We can see that when the quality prediction loss weight $\lambda$ is either too small or too large, the recommendation accuracy decreases, and the recommendation quality is also suboptimal.
In addition, we find that the accuracy loss becomes bigger when $\mu$ is larger, while the quality improvements become marginal when $\mu$ is larger than 0.5.
Thus, we prefer to choose $\lambda=2.0$ and  $\mu=0.5$ to balance recommendation accuracy and quality.

\section{Conclusion}\label{sec:Conclusion}

In this paper, we propose a quality-aware news recommendation method named \textit{QualityRec}, which can recommend higher-quality news to users to improve their experience.
In our approach, we propose a news quality measurement based on the distribution of dwell time.
In addition, we propose a quality-aware user modeling method by using both quality and semantic information to select clicked news for user modeling.
Besides, we propose a quality-aware model training method with an auxiliary quality prediction task to learn quality-aware news models, and a recommendation quality regularization loss to encourage the model to recommend higher-quality news.
Extensive experiments on two real-world datasets show that \textit{QualityRec} can effectively improve recommendation quality and even slightly boost accuracy.
It can not only serve as a recommendation algorithm, but also can be used as a news quality evaluator to help online news platforms to control news qualities.

\bibliography{main}
\bibliographystyle{acl_natbib}
\appendix
\clearpage
\section{Appendix}

\subsection{Experimental Environment}

Our experimental environment is built on a Linux server with Ubuntu 16.04 operation system.
The version of Python is 3.8.
The machine has 4 Tesla V100 GPUs with 32GB memory.
We use Keras 2.2.4 with tensorflow backend 1.15 to implement the models and experiments.

\subsection{Hyperparameter Settings}

The complete hyperparameter settings of our approach are listed in Table~\ref{hyper}.

\begin{table}[h]
\centering
\begin{tabular}{l|c}
\Xhline{1pt}
\multicolumn{1}{c|}{\textbf{Hyperparameters}}& \textbf{Values} \\ \hline
 token embedding dimension                     & 300               \\  
hidden dimension                    & 400   \\ 
attention head                      & 20   \\ 

Transformer layer                     & 1   \\ 
$\lambda$                                 & 2.0         \\
$\mu$                                 & 0.5          \\
dropout                                      & 0.2       \\
max title length                                    & 30  \\
max body length                                    & 200  \\
max history length                                    & 50     \\
optimizer                                    & Adam       \\
learning rate                                 & 1e-4       \\
batch size                                   & 32     \\  
max epoch                                   & 4  \\ 
\Xhline{1pt}
\end{tabular}
\caption{Hyperparameter settings.}\label{hyper}
\end{table}

\subsection{Preprocessing}

Since the quality score of news with few clicks may be noisy, we filter the news with fewer than 10 clicks in the quality prediction task.
In addition, we limit the maximum dwell time to 4096 to filter some extremely long dwell time.

\subsection{Notes on Quality Modeling Method}

To further verify the effectiveness of our proposed quality modeling method, we invite a group of volunteers to rate the quality of a random set of news (the ratings are from 1 to 5).
The results show that the average ratings given by human have a 0.64 PCC score with our quality scores, while only have 0.21 and 0.32 PCC scores with the quality score computing methods (a) and (b) in Section 4.3.  
It shows that our proposed quality modeling method has consistency with the news quality evaluated by human users.

\end{document}